\documentclass[aps,prr,twocolumn,superscriptaddress,longbibliography]{revtex4-2}
\usepackage{graphicx}
\usepackage{bm}% bold math
\usepackage{color}
\usepackage{epstopdf}
\usepackage{amsmath}
\usepackage{url}
\usepackage{braket}
\usepackage{esvect}
\definecolor{darkGreen}{RGB}{0,110,0}
\definecolor{darkBlue}{RGB}{0,0,130}
\usepackage[colorlinks=true, urlcolor=darkBlue,citecolor=darkGreen,linkcolor=darkBlue]{hyperref}
\begin{document}

%Title of paper
\title{Theory of charge-to-spin conversion under quantum confinement}
\author{Alfonso Maiellaro}
\affiliation{CNR-SPIN, c/o Università di Salerno, IT-84084 Fisciano (SA), Italy}
\author{Francesco Romeo}
\affiliation{Dipartimento di Fisica "E.R. Caianiello", Università di Salerno, Via Giovanni Paolo II, 132, I-84084 Fisciano (SA), Italy}
\affiliation{INFN, Sezione di Napoli, Gruppo collegato di Salerno,Italy}
\author{Mattia Trama}
\affiliation{Dipartimento di Fisica "E.R. Caianiello", Università di Salerno, Via Giovanni Paolo II, 132, I-84084 Fisciano (SA), Italy}
\author{Irene Gaiardoni}
\affiliation{Dipartimento di Fisica "E.R. Caianiello", Università di Salerno, Via Giovanni Paolo II, 132, I-84084 Fisciano (SA), Italy}
\author{Jacopo Settino}
\affiliation{Dipartimento di Fisica, Università della Calabria, Via P. Bucci Arcavacata di Rende (CS), Italy}
\author{Claudio Guarcello}
\affiliation{Dipartimento di Fisica "E.R. Caianiello", Università di Salerno, Via Giovanni Paolo II, 132, I-84084 Fisciano (SA), Italy}
\affiliation{INFN, Sezione di Napoli, Gruppo collegato di Salerno,Italy}
\author{Nicolas Bergeal}
\affiliation{Laboratoire de Physique et d’Etude des Mat\'eriaux, ESPCI Paris, Universit\'e PSL, CNRS, Sorbonne Universit\'e, Paris, France.}
\author{Manuel Bibes}
\affiliation{Laboratoire Albert Fert, CNRS, Thales, Universit\'e Paris-Saclay, 91767 Palaiseau, France}
\author{Roberta Citro}
\affiliation{Dipartimento di Fisica "E.R. Caianiello", Università di Salerno, Via Giovanni Paolo II, 132, I-84084 Fisciano (SA), Italy}
\affiliation{INFN, Sezione di Napoli, Gruppo collegato di Salerno,Italy}
\date{\today}

\begin{abstract}
The interplay between spin and charge degrees of freedom in low-dimensional systems is a cornerstone of modern spintronics, where achieving all-electrical control of spin currents is a major goal. Spin-orbit interactions provide a promising mechanism for such control, yet understanding how spin and charge transport emerge from microscopic principles remains a fundamental challenge. Here we develop a spin-dependent scattering matrix approach to describe spin and charge transport in a multiterminal system in the presence of Rashba spin-orbit interaction. Our framework generalizes the B\"uttiker formalism by offering expressions for spin and charge current densities as a function of the lead position, along with the corresponding linear response function. It simultaneously captures the effects of quantum confinement, the response to external magnetic fields, and the  intrinsic properties of the electronic bands, offering a comprehensive description of the spin-charge interconversion mechanisms at play in a Hall bar, in agreement with experiments. 
\end{abstract}
\maketitle
\section{Introduction}
In the last decades spintronics has gained considerable attention due to its potential to revolutionize storage technologies and information processing by exploiting the spin degree of freedom of electrons~\cite{RevModPhys.76.323,Fert,SciPostPhysM,science.1218197,RevModPhys.91.035004}. 
In this field, spin-charge interconversion (SCI) provides a reliable and scalable way to read out spin information electrically, enabling integration into conventional CMOS architectures.
Two fundamental mechanisms enabling SCI in spintronics devices are the Edelstein effect (EE) and the spin Hall effect (SHE). The EE \cite{EDELSTEIN1990233} refers to the generation of a non-equilibrium spin density in response to an applied electric field in systems with strong spin-orbit coupling (SOC), such as Rashba interfaces or topological insulators. This spin accumulation can subsequently diffuse or be injected into adjacent materials, providing a means to manipulate spin currents electrically. The SHE \cite{PhysRevLett.92.126603}, on the other hand, manifests itself as the transverse deflection of spin-polarized carriers in the presence of SOC, leading to the accumulation of spin at the edges of a material and the generation of a transverse spin current. Both effects play a crucial role in the functioning of spin-based electronic devices \cite{PhysRevLett.112.096601,PhysRevB.95.205424,PhysRevB.105.245405,Liu_2007,PhysRevB.108.075418,Min}, and their interplay in confined geometries is subject of intense investigation.\\
Despite extensive theoretical and experimental efforts, a comprehensive understanding of spin transport phenomena in real-space confined systems, particularly in the presence of complex SOC and magnetic textures, is still lacking.\\
\begin{figure}[t!]
	\centering
	\includegraphics[width=0.35\textwidth]{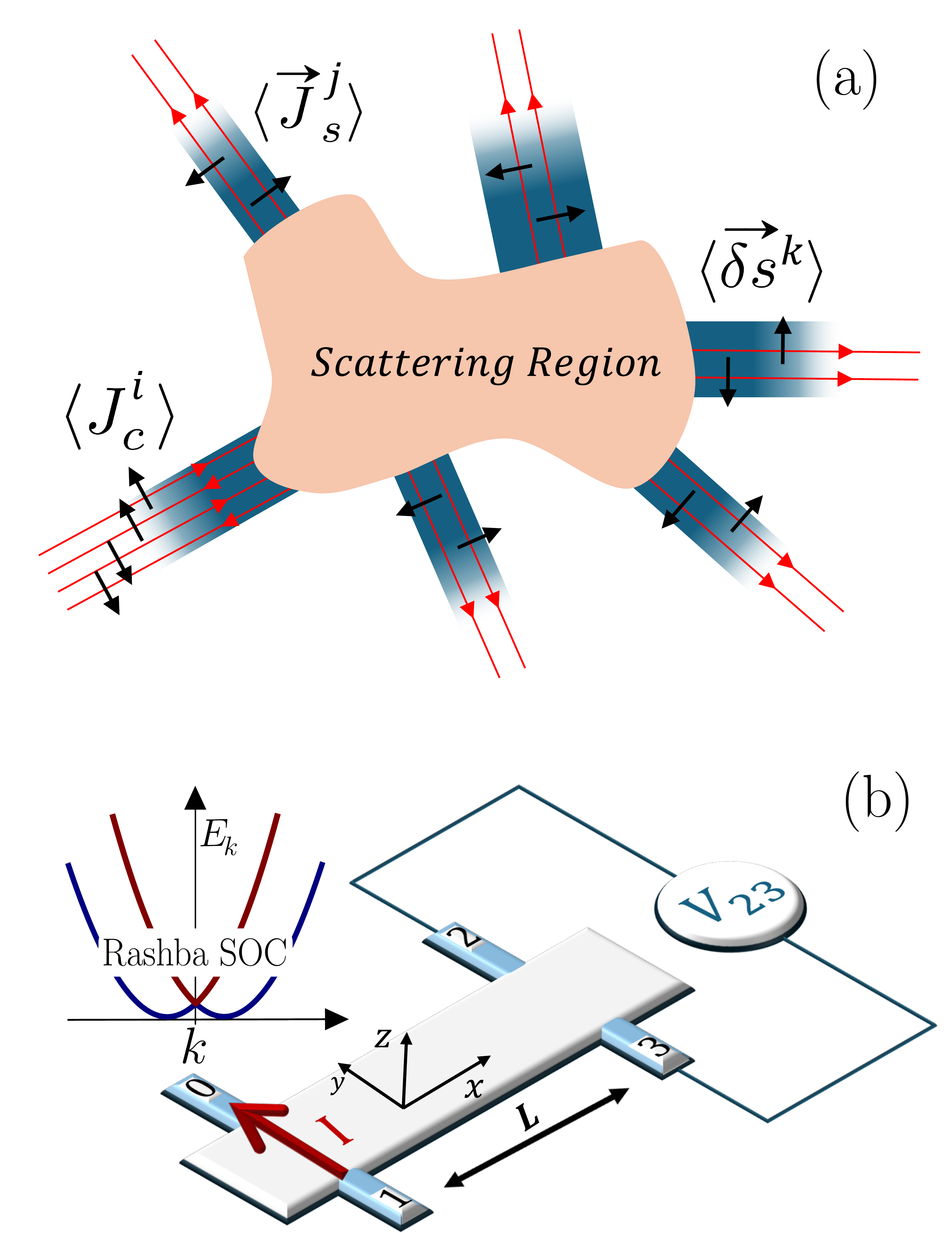}
	\caption{(a) Schematic representation of a multiterminal scattering region through which spin-sensitive transport is measured. The illustration depicts a single scattering event originating from lead $i$, contributing to the charge current $\braket{J_c^i}$, the spin current $\braket{\vec{J}_s^j}$, and the bias-induced spin density $\braket{\vec{\delta s}^{k}}$, in leads $i$, $j$, and $k$, respectively. Spin-resolved incoming and outgoing modes are explicitly indicated by red arrows. (b) Hall bar with Rashba spin–orbit coupling in a current-biased configuration. A net current flows from lead 1 to lead 0. Non-local voltage signal is recorded using leads $2$ and $3$.}
	\label{Figure1}
\end{figure}
A widely used approach to investigate transport in mesoscopic systems is the scattering formalism introduced by B\"uttiker \cite{PhysRevB.46.12485}, which provides a powerful framework to describe  transport in multiterminal geometries. This method has been extensively used to analyze electrical conductance, shot noise, and nonlocal resistance signals in systems with spin-orbit coupling \cite{Romeo,PhysRevB.72.235326,PhysRevB.75.085328,PhysRevB.82.064413}. However, while the B\"uttiker approach excels in capturing charge transport characteristics, it does not inherently provide direct access to quantities such as spin currents, spin densities, or their spatial distributions. These limitations stem from the lack of an explicit and general relation among physical observables~\footnote{A similar distribution of non-local spin currents was obtained in Ref.~\cite{PhysRevB.73.075303} using the bond current approach.}, beyond currents, and the spin-dependent scattering matrix elements. Even though some aspects have been partially addressed in previous works~\cite{Scheid_2007,PhysRevB.73.075303}, a fully developed scattering theory able to analyze spin transport on equal footing with charge transport is necessary to achieve a more complete understanding of SCI mechanisms in mesoscopic systems.\\
%A fully developed scattering theory able to analyze spin transport on equal footing with charge transport is therefore necessary to bridge this gap and to facilitate a more complete understanding of SCI mechanisms in mesoscopic systems.\\
In this work, we develop a theoretical framework providing a non-trivial generalization of the B\"uttiker formalism to explicitly calculate spin currents, spin densities and the Edelstein response and their real-space distributions. By formulating a spin-dependent scattering approach, we provide general expressions for these quantities in multiterminal systems with arbitrary SOC and magnetization profiles. As a typical application of our formalism, we apply it to an H-shaped Hall bar with Rashba SOC, a system known to exhibit non-trivial spin and charge-transport properties. Through our analysis, we uncover the real-space features of the EE and the SHE in the quasi-ballistic regime, demonstrating the dominance of the latter in generating spin currents, in accordance to recent experiments in oxide interfaces \cite{nanolett.9b04079}.
\begin{figure*}
	\centering
	\includegraphics[width=1\textwidth]{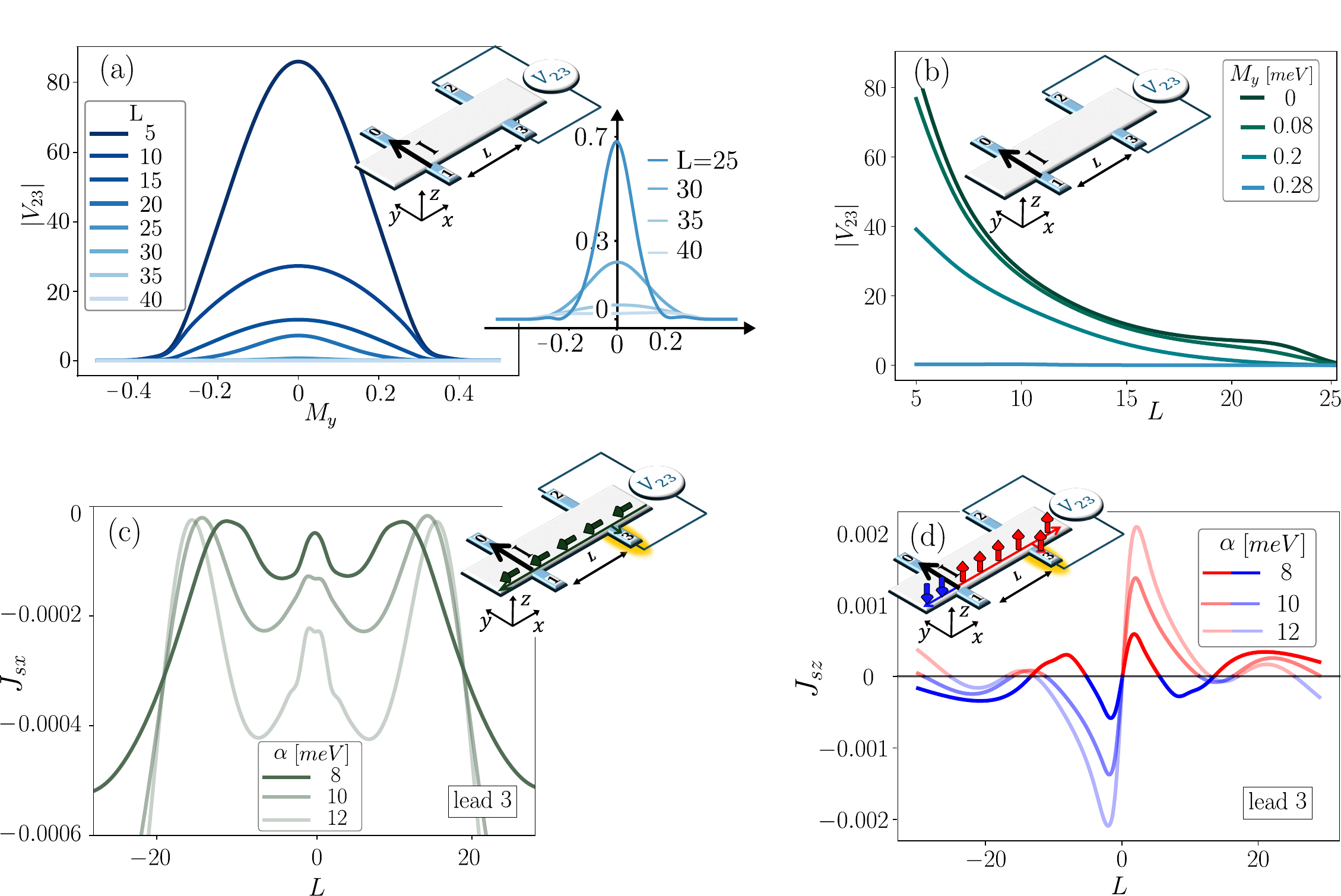}
	\caption{(a) Nonlocal voltage, $|V_{23}|$, as a function of the external magnetic field applied along $y$, $M_y$. The blue gradient corresponds to increasing lead spacings $L$. (b) Vertical slices from panel (a) illustrating the exponential decay of $|V_{23}|$ when the lead spacing is increased. (c) \( x \)-polarized spin current, computed using Eqs.~\ref{Equation1}, \ref{Equation2} for $\vec{M}=0$. The inset illustrates the generation of the \( x \)-polarized spin current via the Edelstein effect. (d) Computed \( z \)-polarized spin current, also obtained from Eqs.~\ref{Equation1}, \ref{Equation2} for $\vec{M}=0$. The inset illustrates the generation of the \( z \)-polarized spin current via the spin-Hall effect.  The source and drain leads are positioned at the center of the system, while the measuring probes are placed either to the left or right. The voltage and spin current curves are expressed in units of \(V_0= 2\pi \hbar I / e^2 \) and \( J_{s}^0=e V_0/4\pi \), respectively, where \( I \) is the applied current bias. For $\alpha = 0$, both $J_s^x$ and $J_s^z$ vanish identically.
	}
	\label{Figure2}
\end{figure*}
\section{Theoretical model}
Building on the scattering matrix approach of Ref.~\cite{PhysRevB.46.12485}, we consider a multimode normal metal lead, modeled as an infinite reservoir, connected to a scattering region, described by an Hamiltonian encoding arbitrary quantum confinement, spin and orbital degrees of freedom. The second-quantized form of the spinful electron field operator in lead $j$, acting on the many-body state and expanded in the transverse modes $m$, is given by: 
\begin{equation}
	\begin{split}
		\hat{\psi}^j(x^j,y^j,t) &= \sum_{m,\sigma} \int_{\mathcal{D}_{jm}} \frac{dE}{\sqrt{\hbar |v^j_{m}(E)|}} e^{-iE t/\hbar}  \ket{\sigma} \times \\ 
		&[\chi_{jm}^+(x^j,y^j) \hat{a}^j_{\sigma m}(E) +\chi_{jm}^-(x^j,y^j) \hat{b}^j_{\sigma m}(E)],
	\end{split}
	\label{EqSCstate}
\end{equation}
where $\hbar$ is the reduced Plank constant and $\ket{\sigma}$, with $\sigma \in \{\uparrow, \downarrow\}$, stands for the eigenstates of the Pauli matrix $\sigma_z$. In writing Eq.~\eqref{EqSCstate}, we assume spin conservation in the leads in order to obtain well-defined spin currents, and that longitudinal and transverse motions are separable, so that the wavefunction of the \( m \)-th transverse mode can be written as $\chi_{jm}^{\pm}(x^j,y^j) = \exp(\pm i k_m x^j) \phi_m(y^j)$, see Appendix~\ref{AppA}. Here, \( x^j \) and \( y^j \) are local coordinates in the reservoir \( j \), with \( x^j \) being the translationally invariant direction of the lead and $y^j$ the transverse direction. Each transverse state \( m \) defines a propagating quantum channel from the reservoir into the scattering region, characterized by the scattering annihilation operator \( \hat{a}^j_{\sigma m}(E) \) and a positive group velocity \( v^j_{m}(E) \), or away from the scattering region, described by \( \hat{b}^j_{\sigma m}(E) \) with a negative group velocity. The wave vector at energy \( E \) is denoted by \( k_m \), and the integration domain \( \mathcal{D}_{jm} \) is restricted to the energy range of the propagating states in the channel \( m \) of the lead \( j \). The total number of quantum channels is \( 2\mathcal{N}^j \), where the factor 2 stems from the spin degeneracy. The incoming and outgoing scattering operators in channel $m$ and lead $j$, \(\hat{a}^j_{\sigma m}\) and \(\hat{b}^j_{\sigma m}\), are related via the scattering matrix \(S\) as $\hat{b}^j_{\sigma m}(E) = \sum_{\sigma' m' j'} S^{jj'}_{\sigma \sigma' m m'}(E) \hat{a}^{j'}_{\sigma' m'}(E)$ to all the channels $m'$ and leads $j'$ appearing in the system. These operators satisfy the relation $\braket{\hat{a}^{j\dagger}_{\sigma m}(E)\hat{a}^{j'}_{\sigma' m'}(E') } \!= \! \delta^{jj'} \delta_{mm'} \delta_{\sigma \sigma'} \delta(E \!- \! E') f^{j'}(E)$, where \(f^{j'}(E)\) is the Fermi-Dirac distribution in lead \(j'\), containing information about the local electrochemical potential and temperature.\\
The expectation value of every physical observable $\mathcal{O}^j$ of the $j$-th reservoir, can be written as $\braket{\mathcal{O}^j}=\int dy^j \braket{\hat{\psi}^{j^\dagger} \mathcal{O}^j \hat{\psi}^j}$, where $\hat{\psi}^j$ is given in Eq. \eqref{EqSCstate}. According to this procedure, detailed in Appendix~\ref{AppA}, we obtain general formulas for the charge current $\langle J_{c}^j \rangle$, spin current $\langle \vv{J}_{s}^j \rangle$, and bias-induced spin density $\langle \vv{\delta  s}^j \rangle$ in the normal metal lead $j$ (see Fig.\ref{Figure1}(a)). Their expressions, which are the central result of this work, when derived at zero temperature and to linear order in $V^j$, take the following form:
\begin{eqnarray}
	\langle J^j_{c} \rangle &=& \!\!\sum_{j'}\!\! \frac{e^2}{2 \pi \hbar} \left[2 \mathcal{N}^j \delta^{jj'}\!\! - \!\!\sum_{m,m'}\!\text{Tr}(S_{mm'}^{jj' \dagger} S_{mm'}^{jj'}) \right] V^{j'}\!\!, \label{Equation1}\\
	\langle \vv{J}^{j}_{s} \rangle &=&\!\! \sum_{j',m,m'} \frac{e}{4 \pi} \text{Tr}(S_{mm'}^{jj' \dagger} \vv{\sigma} S_{mm'}^{jj'}) V^{j'}\!\!, \label{Equation2}\\
	\langle \vv{\delta s}^{j} \rangle &=&\!\! \sum_{j',m,m'} \frac{e}{4 \pi |v^{j}_{m}(\mu)|} \text{Tr}(S_{mm'}^{jj' \dagger} \vv{\sigma} S_{mm'}^{jj'}) V^{j'},\label{Equation3}
\end{eqnarray}
where $\vv{\sigma}$ is the Pauli matrix vector, $e$ is the electron charge, while the mode-dependent group velocity $v^{j}_{m}(\mu)$ characterizes transport at chemical potential $\mu$.\\
The unitarity of the scattering matrix $S$ ($S^\dagger S=I)$, expressed in the form of a tensor product as $S = \sum_{jj' mm' \sigma \sigma'} \mathcal{P}_{jj'} \otimes \mathcal{P}_{mm'} \otimes \mathcal{P}_{\sigma \sigma'}\, S^{jj'}_{mm'\sigma\sigma'}$, where $\mathcal{P}_{\eta \eta'} = \ket{\eta} \bra{\eta'}$ is the projection operator, allows for a direct derivation of probability conservation sum rules involving the mode-resolved scattering amplitudes $S^{jj'}_{mm'\sigma\sigma'}$. These sum rules, which ensure current conservation at the level of individual scattering process, are typically nontrivial to identify within a multimode scattering framework.
Let us also note that the expressions derived in Eqs. (\ref{Equation1})-(\ref{Equation3}) inherently account for the statistical distribution of the leads, explicitly incorporating external driving via the sum over the lead-dependent voltage biases $V^j$. In contrast, a single-process scattering matrix lacks access to this global information and thus cannot capture such experimentally relevant features. Therefore, our formalism overcomes this limitation by enabling a direct theoretical link to experimentally accessible quantities such as spin accumulation and SCI signals. Conceptually, our results within the Büttiker field framework are in the same spirit as the Meir–Wingreen formula \cite{PhysRevLett.68.2512,PhysRevB.50.5528}: while the latter establishes a compact relation between Green’s functions and currents, even in the more general out-of-equilibrium and interacting situation, Eqs. (\ref{Equation1})–(\ref{Equation3}) provide explicit trace-formula expressions that connect the scattering matrix to real-space observables.\\
To implement the tight-binding Hamiltonian of the scattering region, including arbitrary geometries and spin-orbit interactions, we employ the Kwant package \cite{Groth_2014} as an efficient numerical framework to define the system geometry and evaluating its scattering matrix (Fig. \ref{Figure1}(a)).
\begin{figure}
	\centering
	\includegraphics[width=0.45\textwidth]{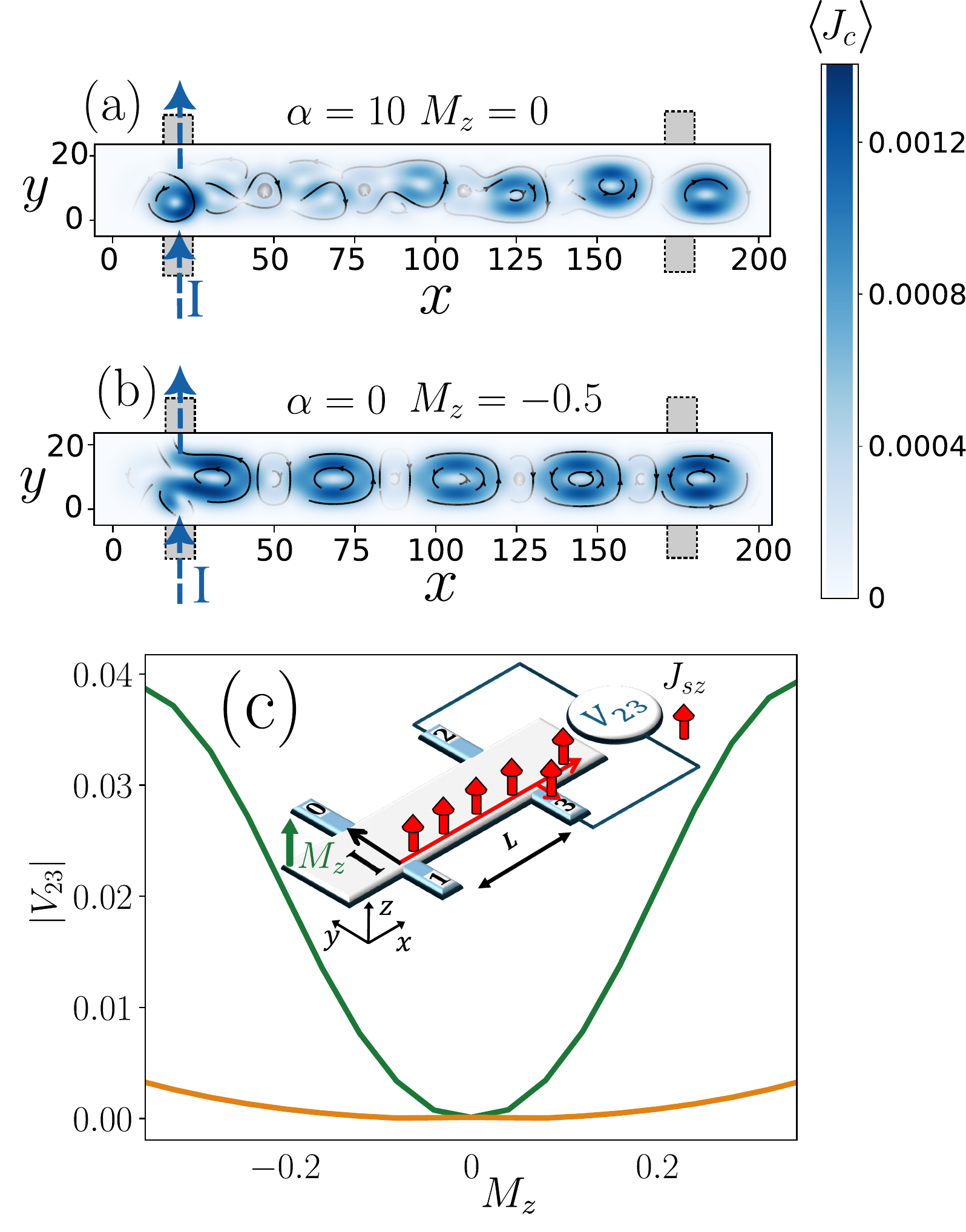}
	\caption{(a), (b) Charge current density in the scattering region for two configurations: (a) $\alpha = 10$, $M_z = 0$; (b) $\alpha = 0$, $M_z = -0.5$. The current is injected from lead 1 and collected at lead 0, with no net current through leads 2 and 3. We set $M_x = M_y = 0$. (c) $|V_{23}|$, expressed in units of \(V_0= 2\pi \hbar I / e^2 \), as a function of the external magnetic field $M_z$. The plot compares results with (green curve) and without (orange curve) the orbital Peierls contribution. The lead distance is fixed to $L = 72$.}
	\label{Figure3}
	%<J_c> in the figure is in mev!!
\end{figure}
While our framework is applicable to both current- and voltage-bias configurations, in this work we focus on the current bias configuration applied to the H-shaped Hall bar, illustrated in Fig.~\ref{Figure1}(b). The Hamiltonian of the scattering region in the continuum limit is given by:
\begin{equation}
\mathcal{H} = \frac{\Pi_x^2+\Pi_y^2}{2\ m_{\rm{eff}}} 
+ \alpha \left( \sigma_x \Pi_y - \sigma_y \Pi_x \right) 
+ \vv{M} \cdot \vv\sigma,
\label{RashbaHcont}
\end{equation}
where \(m_{\rm{eff}}\) is the effective electron mass, \(\alpha\) the Rashba spin-orbit coupling strength, and \(\vv{M} = (M_x, M_y, M_z)\) the Zeeman energy. The generalized momentum \(\vv{\Pi} = \vv{p} + e \vv{A}\) includes the  vector potential \(\vv{A}\) associated to the external magnetic field. To model quantum transport, we discretize Eq.~\eqref{RashbaHcont} on a square lattice, yielding a tight-binding Hamiltonian of a strip of dimensions \(L_S \times W_S\), connected to four semi-infinite, nonmagnetic leads, as shown in Fig.~\ref{Figure1}(b) and detailed in Appendix~\ref{AppB}. In line with recent experiments on oxide interfaces, the parameters are set consistently with previous models for LAO/STO ($001$) \cite{Lesne,Chauleau_2016,Vaz,Trier,PhysRevResearch.3.043170,PhysRevB.103.235120,PhysRevB.107.L201405,PhysRevApplied.22.044012,PhysRevB.109.155306,wójcik2024,Guarcello,PhysRevLett.120.207207,PhysRevLett.121.086801}, thus ensuring realistic energy scales. The comparison between the low-energy spectrum and the electronic structure of the LAO/STO interfaces is reported in Appendix~\ref{AppB}. Let us note that the orbital effect of the magnetic field is incorporated in the tight-binding model via the Peierls phase, which modifies the hopping amplitude as $t 
 \rightarrow t \exp[i (e/\hbar) \int_{\vec{r}_i}^{\vec{r}_j} \vec{A} \cdot d \vec{r} \ ] \equiv t(\vec{r}_i,\vec{r}_j)$, with $t=\hbar^2/(2\ m_{\rm{eff}}\ a^2)$. Moreover, since experiments are close to a quasi-ballistic transport regime \cite{nanolett.9b04079,Ghiasi2019}, we introduce dilute random impurities and perform a statistical averaging of more than 200 disorder realizations (see Appendix~\ref{AppC} for details).\\
 The SCI can be realised in experiments by the two-dimensional spin Hall effect (2D SHE), which leads to a pure spin current transverse to the applied current with an out-of plane spin polarization. The electrical generation
 and detection of this spin current through the direct and inverse 2D SHE have been demonstrated experimentally \cite{Trier}. We thus consider an injected current $I$ from lead 1 to lead 0, i.e., $\langle J_c^j \rangle = (\delta^{1j} - \delta^{0j})I$ and extract the nonlocal voltages at the four terminals to analyze the SCI. We also compute the spin densities and currents via Eqs.~\eqref{Equation2}--\eqref{Equation3} by changing the position of the terminal probe revealing that spin dynamics is consistent with the quasi-ballistic transport regime. 
\section{Numerical results}
Fig.~\ref{Figure2}(a) presents the nonlocal voltage difference between leads 2 and 3, $|V_{23}| = |V_2 - V_3|$, as a function of the Zeeman energy $M_y$. It has a bell-shaped profile and, surprisingly, decays exponentially when the spacing $L$ between the leads increases, as shown in Fig.~\ref{Figure2}(b). This behavior is expected in the diffusive regime~\cite{PhysRevB.79.035304} rather than in the ballistic regime and has been experimentally verified~\cite{nanolett.9b04079}. In fact, due to the spin diffusion length $\lambda_s$ it is experimentally found that $|V_{23}|\propto e^{-L/\lambda_s}$.
This phenomenology is strongly affected by the low-filling regime, i.e. for low values of the chemical potential, where Rashba physics dominates. In contrast, the behavior for higher values of $\mu$ is nontrivial, as multiband physics becomes relevant and additional transport mechanisms emerge~\cite{PhysRevB.66.073311}. Interestingly, when Rashba SOC is present, as illustrated in Appendix~\ref{AppD}, the nonlocal voltage exhibits a slower decay compared to the normal case ($\alpha=0$), revealing the role of SOC in the SCI.\\
Figures~\ref{Figure2}(c) illustrates the computed $x$-polarized spin current $J_{sx}$ at lead 3 as function of its distance from the source/drain leads. This signal originates from the EE~\cite{EDELSTEIN1990233} driven by the electric field along the $y$-direction. Additionally, we observe a $J_{sy}$ component, which is absent in bulk systems. This effect arises from quantum confinement, where the inability of carriers to cross system boundaries induces a tangential electric field that generates an $x$-component of the local field, thereby sustaining $J_{sy}$ via EE.\\
\begin{figure*}
	\centering
	\includegraphics[width=1\linewidth]{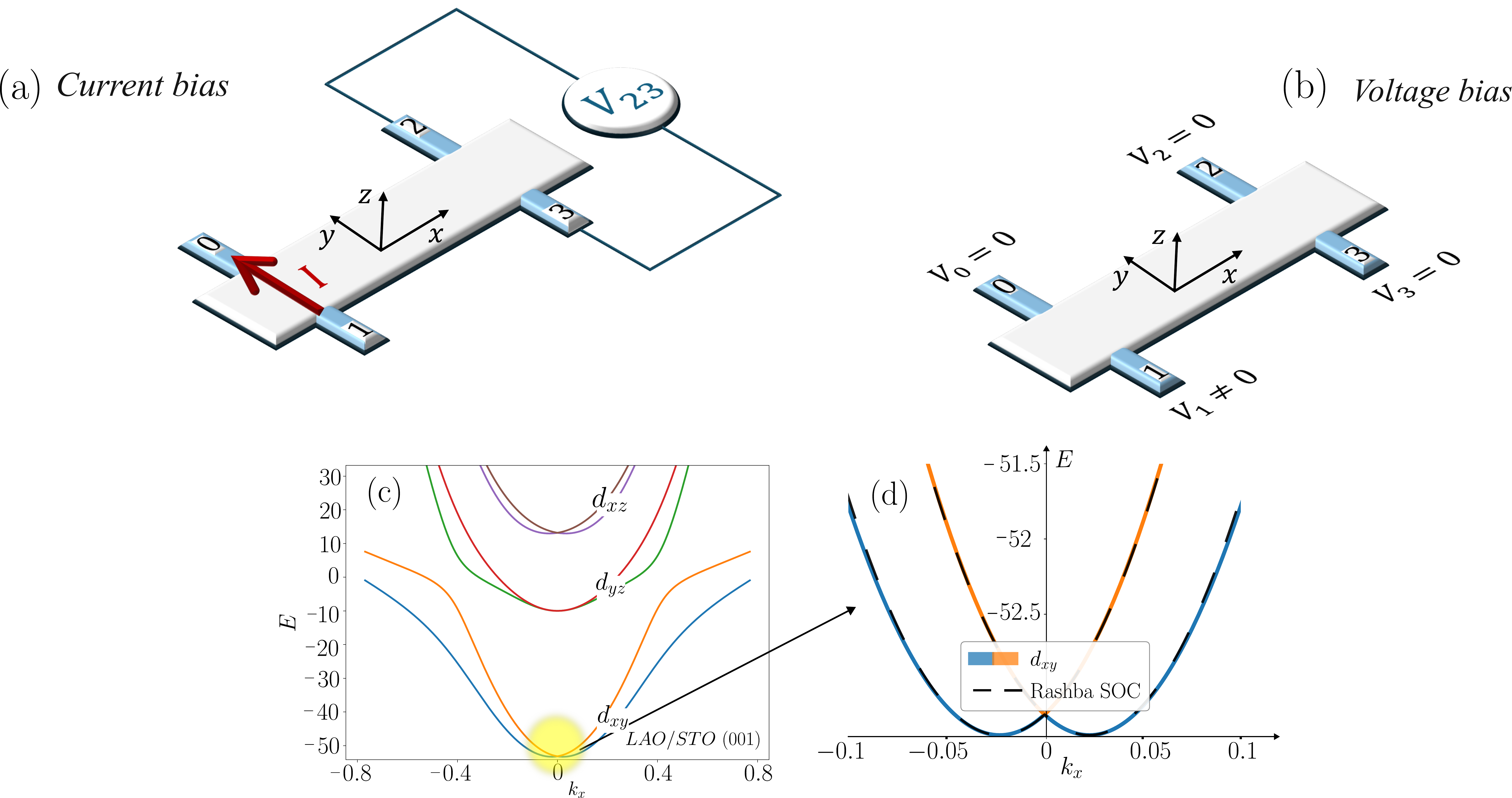}
	\caption{Schematic representation of the oxide Hall bar: (a) net current flux entering from lead $1$ and exiting through lead $0$ (current bias), and (b) fixed bias potentials applied to the four leads (voltage bias). (c) Sketch of the band structure of the LAO/STO $(001)$ interface in one dimension. The six $t_{2g}$ \(d\)-orbitals, i.e., $d_{xy}$-like (blue and orange), $d_{yz}$-like (green and red), and $d_{xz}$-like (purple and brown), are shown. The yellow shaded area highlights the $d_{xy}$ band, matching the Rashba model investigated in the main text when $W_S=1$, as depicted in panel (d).}
	\label{scheme}
\end{figure*}
Our results also encodes the Berry curvature effects, emulating a magnetic field in momentum space \cite{Fuchs2012,Krapek2021}. In fact, the Berry curvature for a Rashba electron gas is expected to generate an effective out-of-plane field that tilts spin polarization along $z$, producing a $J_{sz}$ spin current in which spin-up and spin-down electrons propagate in opposite directions. This effect, depicted in Fig.~\ref{Figure2}(d), is characteristic of the SHE. It can be understood through a gauge field description of the Rashba coupling, which acts as a spin-dependent vector potential associated with the effective out-of-plane magnetic field. Indeed, the gauge field theory, with vector potential $\vv{A} \propto (-\sigma_y, \sigma_x,0)$, introduces conjugate momenta of the form $\Pi_x = p_x - \frac{\alpha}{2}\sigma_y$, $\Pi_y = p_y + \frac{\alpha}{2}\sigma_x$, satisfying the commutation relation $[\Pi_x, \Pi_y] = i\frac{\alpha^2}{2}\sigma_z$. The latter, compared to the commutation relation induced by a perpendicular magnetic field, i.e. $[\Pi_x, \Pi_y] = i\hbar eB$, establishes this equivalence. The sign reversal of $J_{sz}$ as probe $3$ is placed to the left or right of the source/drain leads further supports this interpretation (see Fig. \ref{Figure2}(d)). The residual oscillations in $J_{sx}$ and $J_{sz}$, shown in Figs.~\ref{Figure2}(c) and \ref{Figure2}(d), stem from quantum coherence in the quasi-ballistic regime, being reminiscent of the oscillatory nature of the resonant wave function modes in the scattering region (see Appendix~\ref{AppE}). The spatial distribution of the charge current density, shown in Fig.\ref{Figure3}(a)-(b), provides further insight into the internal transport mechanisms, under conditions where no current flows through leads 2 and 3. In the presence of Rashba SOC ($\alpha \ne 0$, $M_z=0$), circulating current patterns emerge across the scattering region due to resonant mode interference (Fig.\ref{Figure3}(a)). These currents, which develop both $x$ and $y$ components, are responsible for spin accumulation at the lateral contacts. The same phenomenology is found in the complementary case ($\alpha = 0$, $M_z \ne 0$) (Fig.\ref{Figure3}(b)), illustrating the effective equivalence between Rashba SOC and a spin-dependent Zeeman field along the $z$-axis (see Appendix~\ref{AppF}).\\
The Edelstein nature of the in-plane spin currents is also evidenced by reversing the Rashba SOC according to the transformation $\alpha \rightarrow -\alpha$ and observing the validity of the following transformation rules: $J_{s_x}^{\alpha} = -J_{s_x}^{-\alpha}$, $J_{s_y}^{\alpha} = -J_{s_y}^{-\alpha}$ and $J_{s_z}^{\alpha} = J_{s_z}^{-\alpha}$ (details at Appendix~\ref{AppG}). \\
A direct comparison between Figs.~\ref{Figure2}(c) and \ref{Figure2}(d) reveals that the spin current along $z$ exceeds its $x$-component by approximately one order of magnitude, indicating the dominant role of the SHE in this regime. This conclusion is reinforced by the behavior of $|V_{23}|$ under a magnetic field applied along $z$, which exhibits an inverted bell-shaped profile, also known as an inverted Hanle signal (green curve, Fig.~\ref{Figure3}(c))~\cite{Jedema2002,Lou2007}. This characteristic shape arises when the average spin orientation is initially aligned with the external field. Microscopically, it originates from the orbital contribution of the magnetic field, which introduces a Peierls phase in the hopping amplitudes (green curve, Fig.~\ref{Figure3}(c)). Neglecting this orbital contribution leads to a qualitatively different response (orange curve, Fig.~\ref{Figure3}(c)), confirming its crucial role in the observed transport behavior.
\section{Conclusions}
We have developed a scattering matrix approach that provides general expressions for spin and charge transport in multiterminal systems with spin-orbit coupling. This framework extends the B\"uttiker formalism by granting direct access to spin and charge densities and specially response functions, enabling a microscopic analysis of transport phenomena in real space.\\
Applying this formalism to an H-shaped Hall bar with Rashba interaction, we uncover bell-shaped nonlocal voltage profiles that closely resemble those observed in experiments in diffusive regime. Our results provide the first real-space demonstration that these voltage signals originate from charge-to-spin conversion via the Edelstein and spin Hall effects. Additionally, we also uncovered the dominance of the spin Hall effect over the Edelstein effect, as evidenced by an inverted Hanle response under an out-of-plane magnetic field, in agreement with experimental observations.\\
By bridging the gap between ballistic and diffusive regimes, our work establishes a foundation for exploring spin-orbit-driven transport in confined systems, offering new insights into quasi-ballistic spintronic devices and their potential applications in the field of orbitronics.
\section{Acknowledgments.}
The authors acknowledge support from Horizon Europe EIC Pathfinder under the grant IQARO number 101115190. R.C. and F.R. acknowledge funding from Ministero dell’Istruzione, dell’Università e della Ricerca (MIUR) for the PRIN project STIMO(GrantNo. PRIN 2022TWZ9NR). M.B. acknowledge funding from EIC Pathfinder OPEN grant 101129641 (OBELIX). This work received funds from the PNRR MUR project PE0000023-NQSTI (TOPQIN and SPUNTO).
\appendix
\section{Derivation of Charge and Spin Transport Equations}
\label{AppA}
\begin{figure}
	\includegraphics[width=0.45\textwidth]{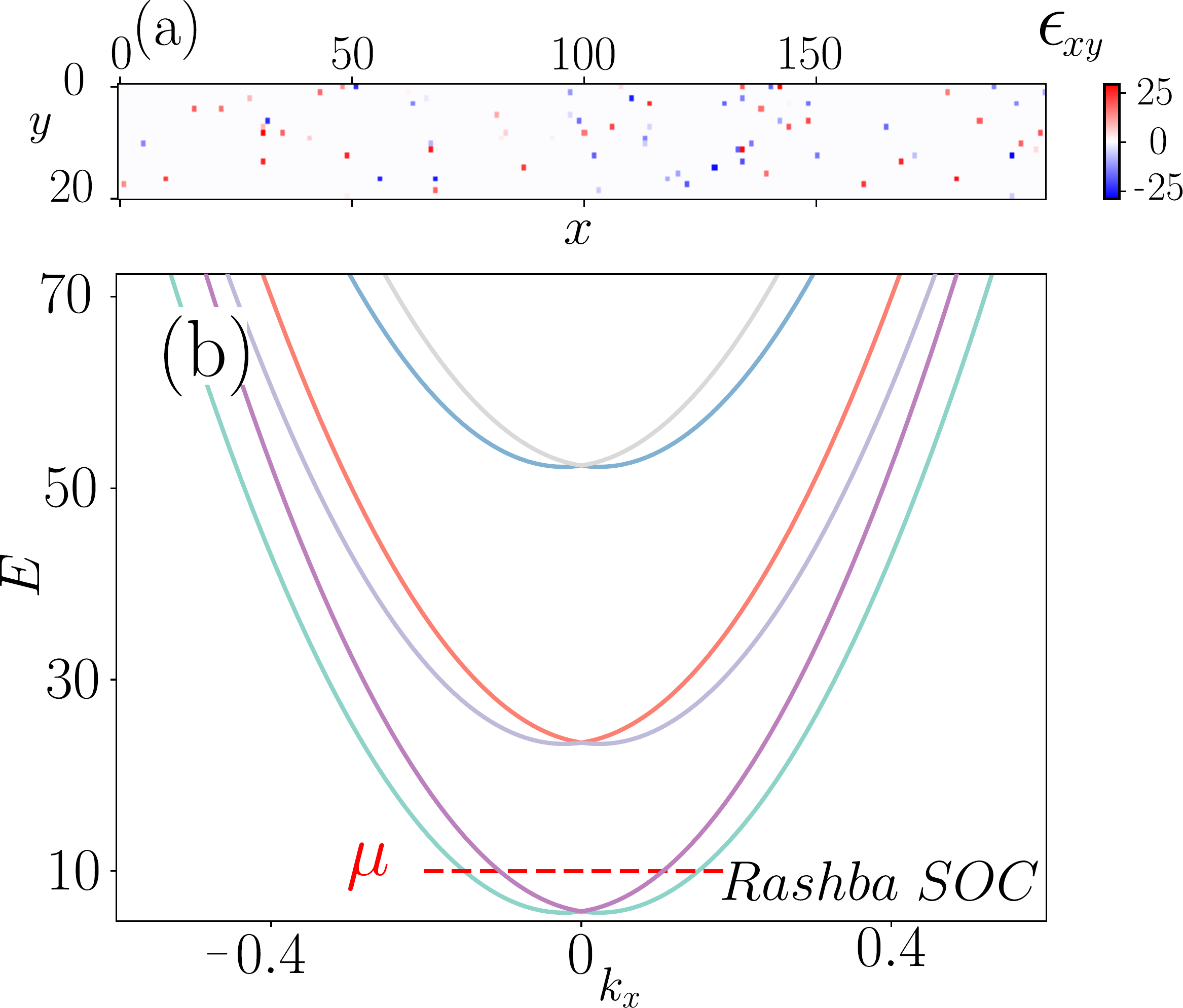}
	\caption{(a) Impurity distribution within a single realization, featuring random potentials distributed with a $2 \%$ density across the strip geometry. (b) The lowest six energy bands of the Rashba SOC model, with the scattering energy $\mu$ explicitly highlighted.} 
	\label{FigSM1}
\end{figure}
In this section, we derive the general expressions for the charge current, spin current, and spin density in a normal metal lead, modeled as an infinite reservoir connected to a scattering region.The Hamiltonian of the lead is given by:
\begin{equation}
	H^{lead} = \frac{p_{x^j}^2 + p_{y^j}^2}{2m} + V(y^j),  \nonumber
\end{equation}
where
\[
V(y^j) =
\begin{cases}
	0, & \text{if } y_0^j - \frac{N_y}{2} \leq y^j \leq y_0^j + \frac{N_y}{2}, \\
	\infty, & \text{otherwise}.
\end{cases}
\]
Here, \( y_0^j \) denotes the center position, and \( N_y \) the transverse width, of lead \( j \). Our derivation follows the scattering approach outlined in Ref.~\cite{PhysRevB.46.12485}. The physical observables can be computed from the second quantized expression of the electron field operator in each lead $j$. Considering spin conserving electrodes, which is a natural choice in order to obtain well-defined spin currents, the field operator can be written as:
\begin{align}
	\begin{split}
		\hat{\psi}^j(x^j,y^j,t) &= \sum_{m,\sigma} \int_{\mathcal{D}_{jm}} \frac{dE}{\sqrt{\hbar |v^j_{m}(E)|}} e^{-iE t/\hbar}  \ket{\sigma} \times \\ 
		&[\chi_{jm}^+(x^j,y^j) \hat{a}^j_{\sigma m}(E) +\chi_{jm}^-(x^j,y^j) \hat{b}^j_{\sigma m}(E)],
	\end{split}
	\label{EqSM1}
\end{align}
where $\ket{\sigma}$, with $\sigma \in \{\uparrow, \downarrow\}$, stands for the eigenstates of the Pauli matrix $\sigma_z$ and 
\begin{equation}
	\chi_{jm}^{\pm}(x^j,y^j)= \sqrt{\frac{2}{N_y}} \sin\left(\frac{m \pi y^j}{N_y}\right) e^{\pm i k_m x^j},
\end{equation}
is the wavefunction of the m-th transverse mode belonging to the $j$ lead having transverse dimension \(N_y\). Here, the wavevector $k_m(E)$ is given by:
\begin{equation}
	k_m(E) = \sqrt{\frac{2m_e E}{\hbar^2} - \left(\frac{m \pi}{N_y}\right)^2},
\end{equation}
with $\hbar$ the reduced Planck constant and $m_e$ the electron effective mass.
In writing Eq. (\ref{EqSM1}) we have introduced the modes group velocity  \(v_{m}^j(E)\) and we have restricted the integration domain \(\mathcal{D}_{jm}\) to the energy range of the propagating states in channel \(m\) of lead \(j\).\\
The incoming and outgoing scattering operators, \(\hat{a}^j_{\sigma m}\) and \(\hat{b}^j_{\sigma m}\), are related via the scattering matrix \(S\) as
\begin{equation}
	\hat{b}^j_{\sigma m}(E) = \sum_{\sigma' m' j'} S^{jj'}_{\sigma \sigma' m m'}(E) \hat{a}^{j'}_{\sigma' m'}(E).
\end{equation}
The scattering operators satisfy the relation
\begin{eqnarray}
	\braket{\hat{a}^{j\dagger}_{\sigma m} (E)\hat{a}^{j'}_{\sigma' m'}(E') } \!= \! \delta^{jj'} \delta_{mm'} \delta_{\sigma \sigma'} \delta(E \!- \! E') f^{j'}(E),\\\nonumber
\end{eqnarray}
where \(f^{j'}(E)\) is the Fermi-Dirac distribution of lead \(j'\).\\
The charge current, spin current, and spin density are defined as the quantum averages of the corresponding operators. Using Eq.~\eqref{EqSM1}, we obtain:
\begin{widetext}
	\begin{align}
		\begin{split}
			\braket{J_c^j} &= \frac{e\hbar}{2 i m_{e}} \int dy^j \Big[ 
			\braket{\hat{\psi}^{j \dagger}(x^j,y^j,t)\ \partial_{x^j} \hat{\psi}^j(x^j,y^j,t)} - \braket{(\partial_{x^j} \hat{\psi}^{j \dagger}(x^j,y^j,t)) \hat{\psi}^{j}(x^j,y^j,t)} \Big] \\ 
			&= \frac{e}{2 \pi \hbar} \sum_{j',m,m'} \int_{\mathcal{D}_{jm}} dE \Bigl[ 2 \delta_{j,j'} \delta_{mm'} - \operatorname{Tr} \Bigl( S^{jj' \dagger}_{mm'} \ S^{jj'}_{mm'}\Bigr) \Bigr] f^{j'}(E).
		\end{split}
		\label{EqSM2}
	\end{align}
	
	\begin{align}
		\begin{split}
			\braket{\vv{J}_s^j} &= \frac{\hbar }{2 i m_{e}} \int dy \Big[ 
			\braket{\hat{\psi}^{j \dagger}(x^j,y^j,t)\ \vv{\sigma}\ \partial_{x^j} \hat{\psi}^j(x^j,y^j,t)} - \braket{(\partial_{x^j} \hat{\psi}^{j \dagger}(x^j,y^j,t))\ \vv{\sigma}\ \hat{\psi}^{j}(x^j,y^j,t)} \Big] \\ 
			&= \frac{1}{4 \pi} \sum_{j',m,m'} \int_{\mathcal{D}_{jm}} dE \operatorname{Tr} \Bigl(S^{jj' \dagger}_{mm'}\ \vv{\sigma}\ S^{jj'}_{mm'} \Bigr) f^{j'}(E).
		\end{split}
		\label{EqSM3}
	\end{align}
	
	\begin{align}
		\begin{split}
			\braket{\vv{s}^j} = \frac{\hbar}{2} \int dy \Big[ 
			\braket{\hat{\psi}^{j \dagger}(x^j,y^j,t) \ \vv{\sigma} \ \hat{\psi}^j(x^j,y^j,t)} \Big] = \sum_{j',m,m'} \frac{1}{4 \pi} \int_{\mathcal{D}_{jm}} \frac{dE}{ |v^j_m(E)|} \operatorname{Tr} \Bigl( S^{jj' \dagger}_{mm'}\ \vv{\sigma}\ S^{jj'}_{mm'} \Bigr) f^{j'}(E),
		\end{split}
		\label{EqSM4}
	\end{align}
\end{widetext}
where $e$ stands for the electron charge, $\vec{\sigma}$ is the Pauli matrices vector and a trace over the spin degrees of freedom is taken. $\partial_{x^j}$ is the derivative with respect to the translational-invariant direction of the lead, being the positive orientation the one entering the scattering region. The above equations are derived in an asymptotic point of the lead, far from the scattering region, where the explicit spatial dependence on $x$ is suppressed. Moreover, as usually done within the scattering formalism, in obtaining Eq. \ref{EqSM4}, we neglected linear contributions in $S$, which are significant only in the vicinity of the scattering region and decay very fast inside the lead. When a DC voltage is applied to the system, the Fermi functions shift as \(f^{j}(E) \rightarrow f(E+eV^{j})\). Thermal transport can be also investigated within the present framework by considering different temperatures $T^j$ in each reservoir. Here, for the sake of simplicity, we set \(T=0\) for every reservoir and to the linear order in the voltage bias, we obtain:
\begin{eqnarray}
	\langle J^j_{c} \rangle &=& \!\!\sum_{j'}\!\! \frac{e^2}{2 \pi \hbar} \left[2 \mathcal{N}^j \delta^{jj'} \!\!- \!\!\sum_{m,m'}\text{Tr}(S_{mm'}^{jj' \dagger} S_{mm'}^{jj'}) \right] V^{j'}\!\!, \label{EqSM5}\\	
	\braket{\vv{J}^{j}_{s}} &=& \sum_{j',m,m'} \frac{e}{4 \pi} Tr\Bigl(S_{mm'}^{jj' \dagger}\ \vv{\sigma}\ S_{mm'}^{jj'}\Bigr) V^{j'}, \label{EqSM6}\\
	\braket{\vv{\delta s}^{j}} &=& \sum_
	{j',m,m'} \frac{e}{4 \pi |v^{j}_{m}(\mu)|} Tr\Bigl(S_{mm'}^{jj' \dagger}\ \vv{\sigma}\ S_{mm'}^{jj'}\Bigr) V^{j'}, \label{EqSM7}
\end{eqnarray}
where \(2\mathcal{N}^j\) is the number of propagating channels in lead \(j\). Eq. (\ref{EqSM7}) represents the term depending on the external voltage, $\braket{\vv{\delta s}^{j}_{s}}=\braket{\vv{s}^j}-\braket{\vv{s}^j}_{eq.}$, where
$$\braket{\vv{s}^j}_{eq.}=\sum_{j',m,m'}\int_{\mathcal{D}_{jm}} \frac{dE \ \theta(\mu-E)}{4 \pi |v^{j}_{m}(E)|} \ Tr(S_{mm'}^{jj' \dagger}\ \vv{\sigma}\ S_{mm'}^{jj'})$$
represents the equilibrium magnetization induced in the $j$ lead due to the proximity with a spin-textured scattering region. Indeed $\braket{\vv{s}^j}_{eq.}$ does not depend on the external voltage and, as a consequence, it does not enter in the charge-spin interconversion mechanisms investigated in this work. In Eq.s (\ref{EqSM5})-(\ref{EqSM7}) the physical observables evaluated at lead $j$ depend on all incident channels $m'$ across every lead $j'$. In particular, once the voltages are assigned on every lead (voltage-bias), the transport quantities can be consistently computed. Equivalently, by reversing Eq. (\ref{EqSM5}) and assigning the current densities of the leads (current-bias), the voltages, spin currents and spin densities can be determined.\\
The above relations are also valid when discrete systems are considered, provided that low-energy regime (i.e. long-wavelength limit) is considered.
\section{Tight-binding Hamiltonian}
\label{AppB}
\begin{figure*}
	\includegraphics[width=1\textwidth]{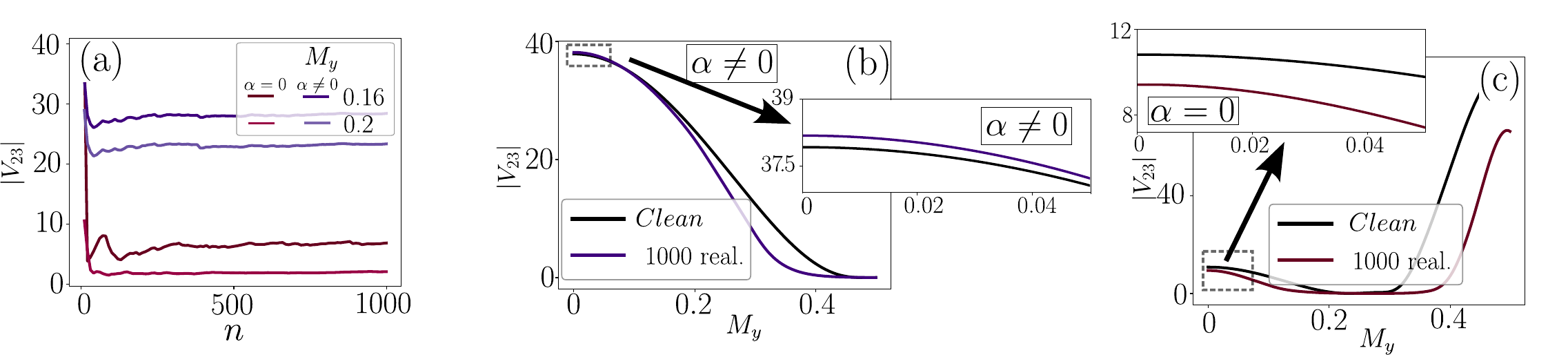}
	\caption{(a) Convergence analysis of the impurity distribution for a density of \( 2\% \), with random on-site energies uniformly distributed in the range \([-30,30]\) meV. The quantity \( |V_{23}| \) is averaged over \( n \) impurity realizations, increasing in steps of $10$ up to \( n = 1000 \), for different values of \( M_y \) and for \( \alpha = 6.026 \) and \( \alpha = 0 \). The curves for \( \alpha = 0 \) are multiplied by a factor of $10$. The $|V_{23}|$ curves discussed in the main text averaging over $1000$ impurity realizations and without averaging are reported in panels (c) for $\alpha=6.026$ and (d) for $\alpha=0$. In all panels, we set $L = 10$, $M_x=M_z=0$.}
	\label{FigSM2}.
\end{figure*}
\begin{figure*}
	\centering
	\includegraphics[width=1\textwidth]{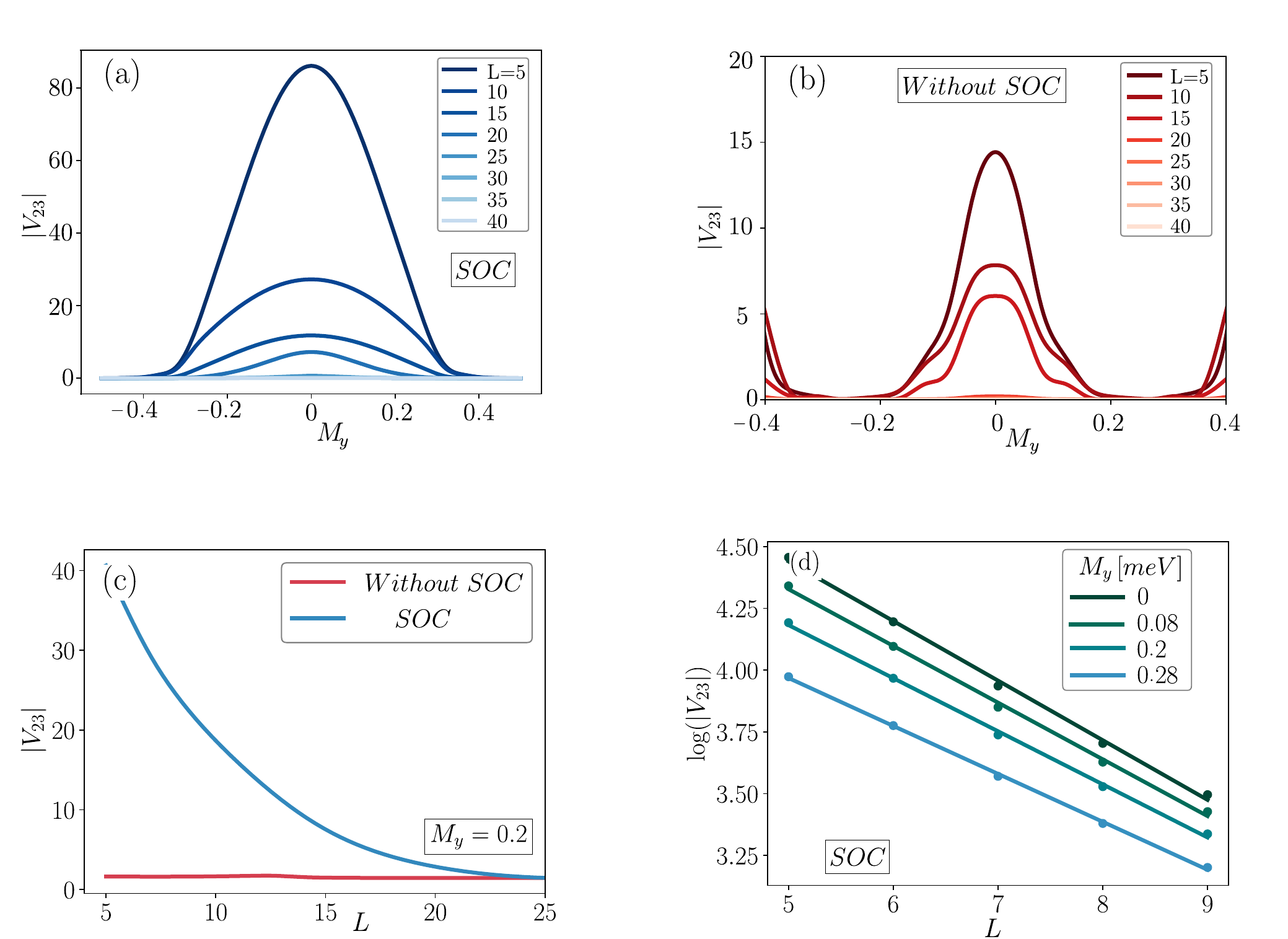}
	\caption{(a) Nonlocal voltage \( |V_{23}| \) as a function of \( M_y \) in the presence of Rashba SOC. The figure is taken from the main text.  
		(b) Nonlocal voltage \( |V_{23}| \) as a function of \( M_y \) in the absence of Rashba SOC.  
		(c) Evolution of \( V_{23} \) as the lead distance \( L \) increases, comparing cases with and without SOC. (d) The same plot as in Fig. 2(b) of the main text but in log-scale. The interpolating lines for inceasing $M_y$ are described by the equations $y \simeq-0.24x+5.65$, $y\simeq-0.23x+5.47$, $y\simeq-0.21x+5.26$, $y\simeq-0.19x+4.94$. In all panels, we set $M_x=M_z=0$.}
	\label{FigSM3}
\end{figure*}
The Hamiltonian of the Rashba model can be described as $H = H_0 + H_{SO} + H_M$, with:
\begin{eqnarray}
	\label{HamKinetic}
	\begin{split}
		H_0\!=\! &\sum_{x,y} \Bigl[ (-\epsilon+4t) \Psi^{\dagger}_{x,y} \sigma_0 \Psi_{x,y} -t_1 \Bigl(\Psi^{\dagger}_{x,y} \sigma_0 \Psi_{x+1,y} + H.c\Bigr)+\\
		&-t_2 \Bigl(\Psi^{\dagger}_{x,y} \sigma_0 \Psi_{x,y+1}+ H.c.\Bigr) \Bigr] 
	\end{split}
	\label{HamK}
\end{eqnarray}
\begin{eqnarray}
	\label{Rashba}
	H_{SO}\!=\! \mathrm{i} \alpha \Bigl(\!\sum_{x,y}\! \Psi^{\dagger}_{x,y} \sigma_y \Psi_{x+1,y}\!\! -\!\Psi^{\dagger}_{x,y} \sigma_x \Psi_{x,y+1}\!\Bigr)\!\!+\!\! H.c.
	\label{HamSOC}
\end{eqnarray}
\begin{eqnarray}
	\label{Zeeman}
	H_{M}= \sum_{x,y} \Psi^{\dagger}_{x,y}\ \vv{M} \cdot \vv{\sigma}\ \Psi_{x,y}.
	\label{HamZeeman} 
\end{eqnarray}	
The terms $H_0$, $H_{SO}$, and $H_M$ in the Hamiltonian correspond to the kinetic energy, spin-orbit coupling, and Zeeman interaction, respectively. Specifically, $\epsilon$, $t$, $\alpha$, and $\vv{M} = (M_x, M_y, M_z)$ denote the on-site energy, hopping amplitude, Rashba spin-orbit coupling, and Zeeman energy. The presence of the vector potential associated with the Zeeman term $M_z$ introduces Peierls phases in the hopping amplitudes, modifying them as  $t_1 = t e^{\frac{iM_z}{2tg}y}$, $t_2 = t e^{\frac{-iM_z}{2tg}x}$, where $g$ is the Landé g-factor. Since the system is confined to the $x$-$y$ plane, only the $z$-component of the magnetic field contributes to the Peierls phases. $\Psi_{xy} =(c_{xy, \uparrow}\ c_{xy,\downarrow})^T$ denotes a vector whose components are the electron annihilation operators for a given spin and position. The summation over indices $(x,y)$ spans over the length $L_S$ and the width $W_S$ of the strip. $\sigma_0$ indicates the identity matrix.\\
In order to set realistic parameters we consider an effective model for (001) interface of bulk LAO/STO, a promising platform for the investigation of charge-spin interconversion mechanisms, see Fig. \ref{scheme}. According to Ref.~\cite{PhysRevB.103.235120}, in the low-energy limit and for small wave vectors \( k \), the \( xy \)-like band of LAO/STO can be effectively described by a Rashba Hamiltonian. In particular, the hopping amplitude \( t \) and the Rashba coupling parameter \( \alpha \) in our system can be expressed in terms of the effective electron mass \( m_{eff} = 0.8\, m_e \), the lattice constant \( a = 0.39 \) nm, the atomic spin-orbit coupling \( \Delta_{so} = 10 \) meV, the inversion symmetry breaking coupling \( \gamma = 40 \) meV, and the tetragonal crystal field potential \( \Delta_t = -50 \) meV at the (001) interface of bulk LAO/STO. Specifically, these parameters are given by the relations  
\begin{equation}  
	t = \frac{\hbar^2}{2m_{eff}a^2} \simeq 313\ \text{meV},
\end{equation}
\begin{equation}	
	\alpha = \frac{\sqrt{2}\ \gamma\ \exp \left( \operatorname{arcsinh} \eta_R \right)}{2\ [1 + \exp \left( 2 \operatorname{arcsinh} \eta_R \right)]}\simeq 6.026\ \text{meV},  
\end{equation}  
where \(\eta_R=\sqrt{2}(|\Delta_t|+\Delta_{so})/4 \Delta_{so}\). Throughout this work, we assume \( \epsilon = 0 \), \( L_S = 200 \), and \( W_S = 20 \), while considering an extended \( \alpha \) range to encompass the physics of devices fabricated with different materials.
\section{From ballistic to quasi-ballistic regime}
\label{AppC}
\begin{figure}
	\includegraphics[width=0.45\textwidth]{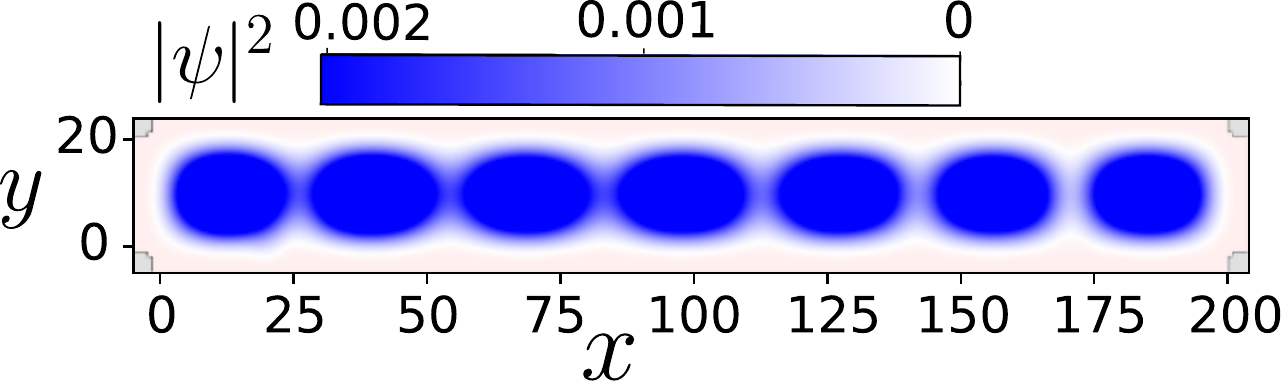}
	\caption{Probability density along the scattering region for a single scattering event, where an electron is incoming from lead 1 with \( L = 30 \). The remaining parameters have been set as $M_x=M_z=0$, $M_y=0.28$, $\mu=10$, $\alpha=6.026$.}
	\label{FigSM4}
\end{figure}
\begin{figure*}
	\includegraphics[width=1\textwidth]{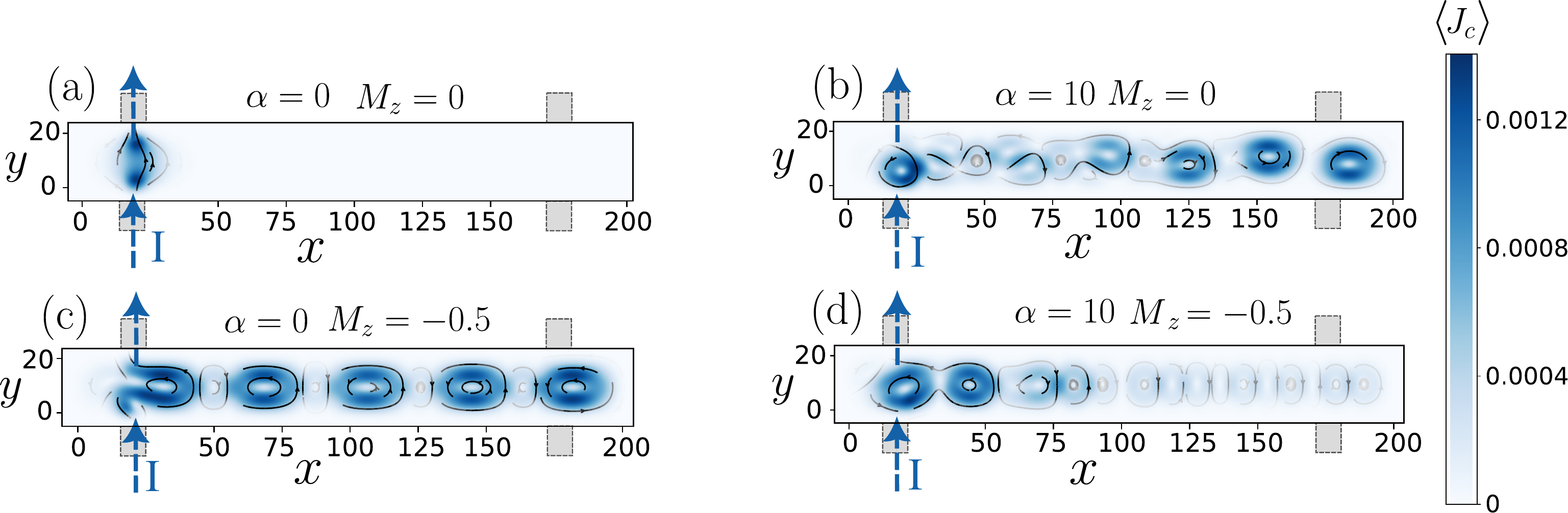}
	\caption{Charge current density along the scattering region for different values of $\alpha$ and $M_z$. The current is injected from lead $1$ and extracted from lead $0$, while no current enters or exits through leads $2$ and $3$. Leads $0$ and $1$ are attached at $x=20$ of the scattering region, while $M_x=M_y=0$.}
	\label{FigSM5}
\end{figure*}
Trough the strip we add a random strength potential $\epsilon_{xy} \in [-30,\ 30]\ meV$ randomly distributed over the strip with a $2 \%$ density. A single impurity realization is reported in Fig. \ref{FigSM1}(a). In Fig. \ref{FigSM1}(b) we show the lowest six energy bands of the Hamiltonian, with the chemical potential $\mu$ explicitly highlighted.
Incorporating disorder in the scattering region is a necessary ingredient to account for quasi-ballistic regime induced by random electron-impurity interactions. In our numerical simulations we consider up to 1000 impurity realizations. In Fig. \ref{FigSM2}(a), we show the nonlocal voltage as a function of the number of impurity realizations, $n$, obtained by considering different Zeeman energies, both with and without Rashba SOC, respectively. For small values of $M_y$ and in the presence of SOC, impurity scattering enhances the nonlocal signal, as shown in panel (b), whereas in the absence of SOC, disorder reduces the voltage signal $|V_{23}|$, panel (c), thereby reinforcing the key findings presented in the main text. Based on this analysis and in order to optimize the numerical procedure, we set the compromise value $n=200$ for the disorder realizations used to obtain the main text curves. Indeed, at this critical threshold convergence is almost reached compared to simulations performed using $1000$ disorder realizations.
\begin{figure*}
	\includegraphics[width=1\textwidth]{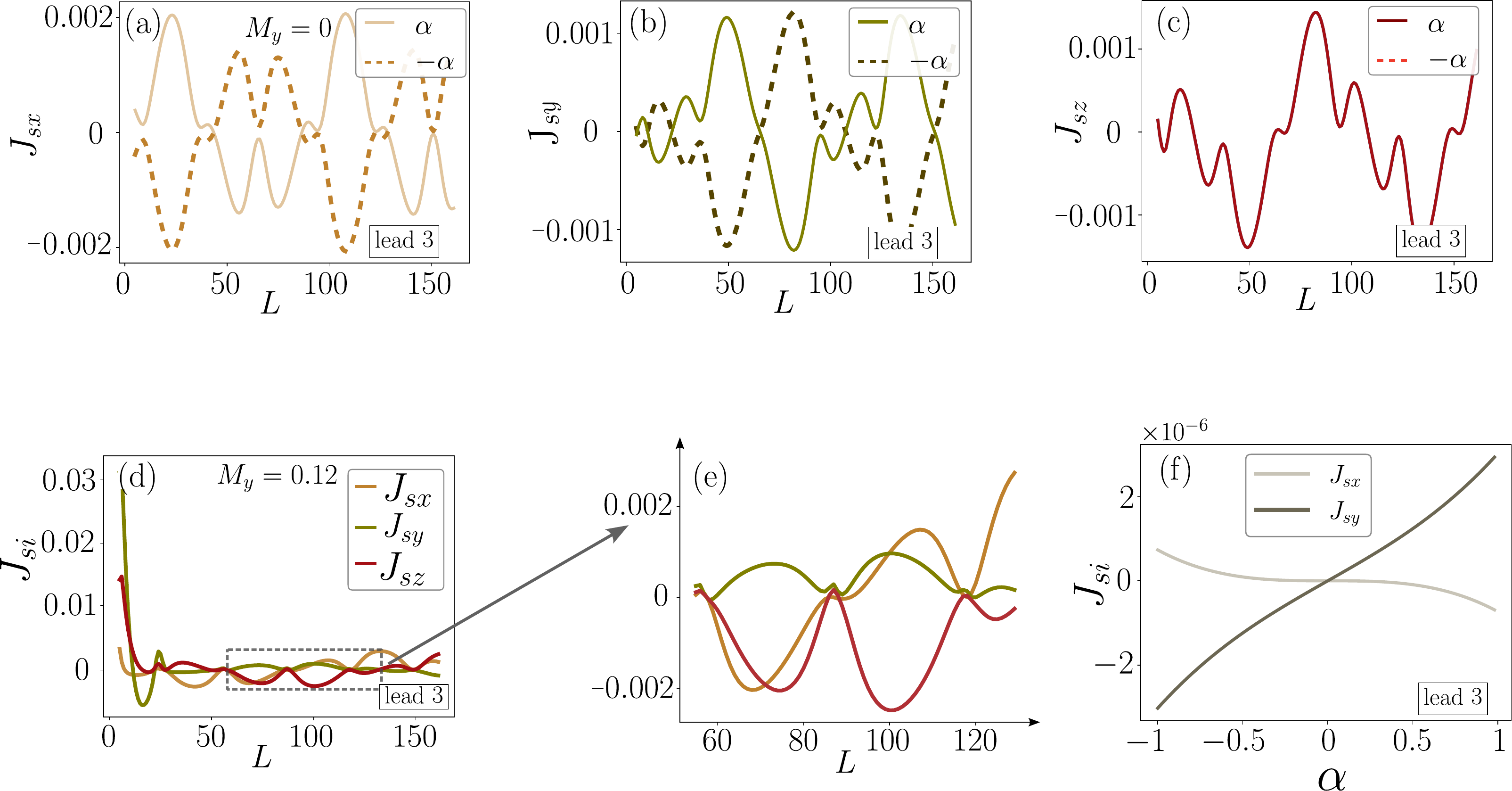}
	\caption{(a)-(c) Spin currents as a function of $L$ at zero Zeeman field, $\vec{M}=(0,0,0)$, with the two orientations of the Rashba SOC axis. Panels (a), (b), and (c) show the $x-$, $y-$, and $z$-components, respectively. (d) Oscillatory behavior of spin current components vs $L$ for $M_y=0.12$. The zoomed region in (e) highlights the weak oscillations of the $y$-component respect the other components. We set $\alpha=12$ for panels (a), (b) and (c) and $\alpha=6.026$ for panels (d) and (e). (f) $J_{sx}$ and $J_{sy}$ as a function of $\alpha$ for $L=10$ and $\vec{M}=(0,0,0)$.}
	\label{FigSM6}
\end{figure*}
\section{Effect of Rashba SOC on nonlocal voltages}
\label{AppD}
In Fig.~\ref{FigSM3}, we investigate the influence of Rashba spin-orbit coupling (SOC) on the nonlocal voltage response in the presence of an in-plane Zeeman field \( M_y \). The comparison between scenarios with \( \alpha = 0 \) and \( \alpha \neq 0 \) reveals a substantial modification in the transport behavior along the Hall bar. Specifically, when \( \alpha = 0 \), the nonlocal voltages measured between leads 2 and 3 are significantly reduced compared to the case with finite SOC, as illustrated in Fig.~\ref{FigSM3}(a)-(b). Furthermore, the voltage decay with increasing lead distance \( L \) is markedly steeper in the absence of Rashba SOC, as evidenced by Fig.~\ref{FigSM3}(c). This observation suggests that SOC not only enhances the magnitude of nonlocal signals but also extends their spatial coherence, mitigating the rapid attenuation induced by the Zeeman field alone.
To quantitatively characterize this behavior, Fig.~\ref{FigSM3}(d) presents the exponential decay of \( |V_{23}| \), in presence of SOC, as a function of \( L \). The continuous lines represent the linear fits performed in the logarithmic scale, highlighting the exponential nature of the decay process, whose mathematical expressions are provided in the caption of the figure. 

\section{Local densities for a single scattering process}
\label{AppE}
In Fig.~\ref{FigSM4}  we analyze the probability density  for a single scattering event. An electron is injected from lead 1 with $L = 30$ in the presence of a strong external field $M_y = 0.28$. The probability density of the resonant mode extends throughout the entire system, exhibiting spatial oscillations, with the same nodal structure appearing in the transport observables of the main text.
\section{Charge Current Density in the Scattering Region}
\label{AppF}
In Fig. \ref{FigSM5}, we analyze the charge current density along the scattering region in the absence of \(M_z\), both without and with \(\alpha\), shown in panels (a) and (b), respectively. Similarly, panels (c) and (d) correspond to the case \(M_z=-0.5\), again without and with \(\alpha\). A single electron is injected from lead 1 into lead 0, while no current flows through leads 2 and 3, as clearly illustrated in Fig. \ref{FigSM5}(a). When \(\alpha \neq 0\) (Fig. \ref{FigSM5}(b)), the internal electric field of the scattering region develops a complex structure, leading to the emergence of both \(x\) and \(y\)-components in the charge current density. As a result, spin current components also accumulate inside the Hall bar due to the Edelstein effect, and are consequently observed in the non-local leads 2 and 3, as discussed in the main text. Furthermore, the vortex-like structure, which extends across the entire sample, is reminiscent of the nodal patterns previously identified in Fig. \ref{FigSM4}. A similar behavior is observed for \(M_z \neq 0\) with \(\alpha = 0\), suggesting that Rashba coupling effectively acts as a spin-dependent Zeeman field along the \(z\)-axis. This can be demonstrated via a gauge field description of Rashba coupling. The standard Rashba Hamiltonian can be rewritten as:
$$\tilde{H} = \frac{\Pi_x^2 + \Pi_y^2}{2m} - \frac{\alpha^2}{4m},$$
with the modified momenta defined as $\Pi_x = p_x - \frac{\alpha}{2}\sigma_y$, $\Pi_y = p_y + \frac{\alpha}{2}\sigma_x$.
These satisfy the commutation relation:
\[
[\Pi_x, \Pi_y] = i\frac{\alpha^2}{2}\sigma_z.
\]
In comparison, for a magnetic field \( \vv{B} = (0, 0, B) \), the Hamiltonian is: $\tilde{H} = \frac{\Pi_x^2 + \Pi_y^2}{2m}$,
with momenta $\Pi_x = p_x + eBy$, $\Pi_y = p_y$, leading to the commutation relation:
\[
[\Pi_x, \Pi_y] = i\hbar eB.
\]
By comparing these results, we notice that the Rashba interaction introduces a spin-dependent effective magnetic field along \(z\), with field strength of $\frac{\alpha^2}{2e\hbar}$ for spin-up electrons and $-\frac{\alpha^2}{2e\hbar}$ for spin-down electrons.\\
In Fig. \ref{FigSM5}(d), when both \(M_z\) and \(\alpha\) are present, we notice that an appropriate choice of their relative signs can lead to a suppression of the charge current density along the scattering region.

\section{Further results on the spin currents}
\label{AppG}
In Figs.~\ref{FigSM6}(a)-(c), we present the $x$, $y$, and $z$ components of the spin current density when the Rashba SOC axis is reversed. We observe that $J_{s_x}$ and $J_{s_y}$ invert their sign under the transformation $\alpha \rightarrow -\alpha$, while $J_{s_z}$ remains unchanged. This behavior is consistent with the charge-spin interconversion mechanisms described in this work. Indeed, the spinorial wave function $\varphi^{\alpha}$ with Rashba coupling $\alpha$ is related to the wave function with the reversed Rashba axis, $\varphi^{-\alpha}$, through the unitary transformation $\varphi^{\alpha} = \hat{\sigma}_z \varphi^{-\alpha}$. The spin current is given by  
\[
\vv{J}_{s}^{\boldsymbol{\alpha}} \propto -\Im \Bigl[ \varphi^{\alpha \dagger} \vv{\sigma} \bigl( \boldsymbol{\nabla} \varphi^{\alpha} \bigr) \Bigr],
\]  
where $\Im$ is the imaginary part and $\boldsymbol{\nabla}$ is the gradient operator. By combining these relations and recalling that $\hat{\sigma}_z \hat{\sigma}_i \hat{\sigma}_z = (2 \delta_{zi} - 1) \hat{\sigma}_i$, we obtain the symmetry relations $J_{s_x}^{\alpha} = -J_{s_x}^{-\alpha}$, $J_{s_y}^{\alpha} = -J_{s_y}^{-\alpha}$, and $J_{s_z}^{\alpha} = J_{s_z}^{-\alpha}$.\\
In Fig.~\ref{FigSM6}(d), we show the oscillatory behavior of the spin current as the system length \(L\) increases for \(M_y = 0.12\). These oscillations originate from the nodal structure of the probability density, as reported in Fig.~\ref{FigSM4}. The inset in Fig.~\ref{FigSM6}(e) emphasizes that the amplitude of oscillations in the \(y\)-component of the spin current is smaller compared to those observed in the remaining components.
Finally, in Fig.~\ref{FigSM6}(f), we present the dependence of \(J_{s_x}\) and \(J_{s_y}\) on \(\alpha\). These spin currents are induced by the Edelstein effect and exhibit a linear scaling with \(\alpha\) in the small \(\alpha\) regime.
\end{document}